# Hostile Intent Enumeration using Soft Computing Techniques

Souham Biswas[1], Manisha J. Nene[2]

[1] J.K. Institute of Applied Physics & Technology, University of Allahabad
Allahabad, 211002, India
**souhambiswas@outlook.com**

[2] Dept. of Applied Mathematics and Computer Engineering, Defence Institute of Advanced Technology, Defence R&D
Organization, Ministry of Defence, Govt. of India
Pune, 411025, India
**mjnene@diat.ac.in**

**Abstract**
In any tactical scenario, the successful quantification and triangulation of potential hostile elements is instrumental to minimize any casualties which might be incurred. The most commonly deployed infrastructures to cater to this have mostly been surveillance systems which only extract some data pertaining to the targets of interest in the area of observation and convey the information to the human operators. Accordingly, with the ever increasing rate at which warfare tactics are evolving, there has been a growing need for "smarter" solutions to this problem of hostile intent enumeration. Recently, a number of developments have been made to ameliorate the efficacy and the certitude with which this task is performed.
This paper discusses two of the most prominent approaches which address this problem and posits the outline of a novel solution which seeks to address the shortcomings faced by the existing approaches.

## 1. Introduction

The modern age has witnessed a significant burgeoning of attack/defence tactics and keeping with the pace, significant work has also been carried out in the field of hostility detection. The problem of hostile intent detection basically seeks to *numerically quantify* the trait of hostility. The biggest hurdle faced in this process is the inherently ambiguous nature of this attribute; hostility has different meanings for different observers.

The human notion of hostility derives from the characteristic of *intuition* which is hardwired in human beings. But, to incorporate a similar functionality into a machine, a number of variables need to be taken into account. The computational complexity needed to process these variables analytically is often astronomical. But disregarding the predicament posed by this complexity, a number of approaches have been made to address it.

Most of the present solutions catering to this problem statement of hostility determination are purely analytical in nature and hence conform to a mathematically predefined notion of "hostility". These approaches provide the methodologies and algorithms instrumental to their implementation in the software realm. This paper enumerates a set of such soft computing solutions which are prominent in this field, and posits the outline of a novel approach to this problem which makes for the shortcomings faced in the current approaches.

## 2. Solutions in Deployment

Hostility of an object can be defined as a quantity whose magnitude is symbolic of the probability that the object will commit actions in interests that conflicts with that of the observer.

Terminologies used:

- **Hostility:** The quantified degree which denotes the potential probability of hostile behavior as defined above.
- **Area of interest**: This refers to the field of vision which encompasses the objects being monitored for hostile traits.
- **Object**: This refers to a moving entity which is subject to probation for the evaluation of its hostility.

Enumerated in this section are the analyses of some of such popular techniques.

2.1 Advanced Surveillance

Advanced Surveillance [1] works in conjunction with human expertise to achieve results. It has been implemented as a software solution which utilizes internet services to process the data pertaining to the location of various objects in the given area of interest by querying large databases which store corresponding information crucial to its operation. As the name suggests, the basic function performed by this approach is basically surveillance. It does not explicitly provide a measure of hostility of any object in the area of interest, for that decision is left to the human agent manning the system.

This system comprises of a display showing the various objects inside a given field of interest, with human agents continuously monitoring the same. It has an edge over traditional surveillance systems like radars, as it works through the deployment of "agents". Agents are basically elements which search for certain data traits in the databases to which it is linked (commercial or government databases) or the incoming data stream and enumerate those, which conform to a set of predicates which can be defined by the operator. Various predefined templates are available to the operator for performing the operation. Here, templates refer to sets of different predicates for specialized purposes.

For example, if the operator selects the template for speed violations, an agent is launched which analyses all the data and highlights those objects which are in violation with the given constraints; in this case, speed. The data being analyzed is drawn from a variety of commercial and military databases. The system also provides provisions for extracting significant amount of information pertaining to a single object inside the area of interest such as, its previous locations within a given time frame.

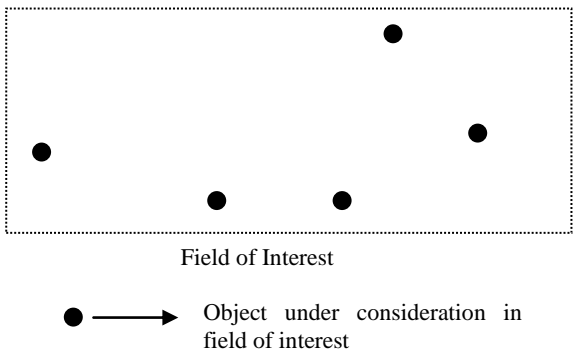

Fig. 1 Objects in Field of Interest, as viewed on a Traditional Surveillance System.

Shown in Fig.1 is a representation of what one would typically expect to see on the display of a surveillance system which does not have the aforementioned advanced capabilities. As we can see, the operator is provided with a view of only the present scenario pertaining to the locations of the objects. The efficacy of this system depends largely on the ability of the human operator to discern with certitude, the hostility of the various objects presented to him. Therefore, this system has a larger potential to yield dicey results and false alerts as the human element has a considerable role to play in it.

In contrast to this, the working of the advanced surveillance system is much different and is illustrated using Fig.2 and Fig. 3.

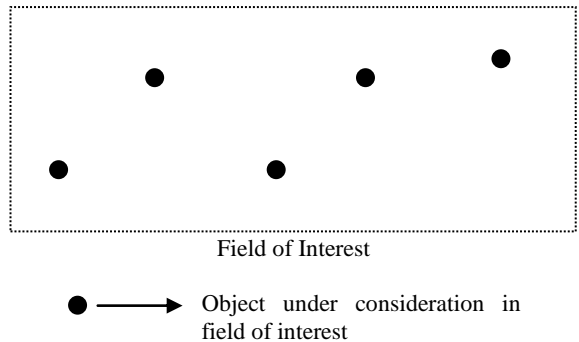

Fig. 2 Objects in Field of Interest, as viewed on the Advanced Surveillance System with no Template Selected or Agent Deployed.

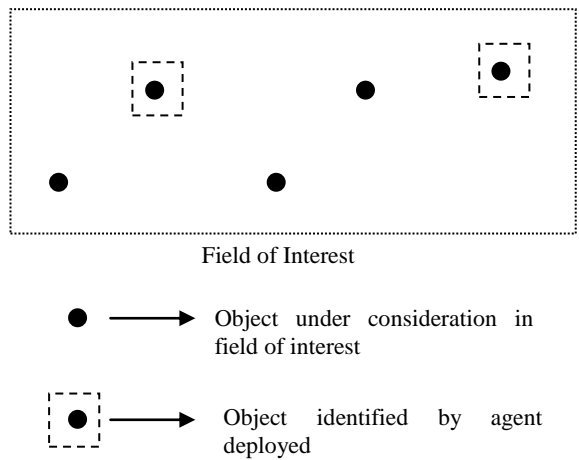

Fig. 3 Objects in Field of Interest, as viewed on the Advanced Surveillance System with some Template Selected or Agent Deployed..

As seen in Fig. 2, it depicts an illustration of what is viewed by the human agent when no template is selected or an agent is deployed. This resembles the output of the traditional surveillance system. But, after the operator deploys an agent by selecting, for example a template for speed violation, specific objects in the field of interest are explicitly highlighted as illustrated in Fig. 3. This process is carried out by the deployed agents which access a variety of databases, and get information pertaining to the objects. Subsequently, it is upto the operator to finally decide as to which object could be classified as potentially hostile.

Therefore as we can see, the data processing tasks which were done manually, have been extensively automated. Although this system offloads a lot of strain from the human operator, it cannot be deemed as a foolproof solution to the problem of hostile intent enumeration as it does have provisions for human error. Therefore, a fully automated solution for the same is highly desirable.

## 2.2 Detection of Hostile Intent from Movement Patterns

Detection of Hostile Intent from Movement Patterns [2] overcomes the shortcoming of the previous approach by automating the entire process and hence relieving the human agent of the responsibility held previously. This method discretely follows a set of given instructions and performs deterministic mathematical calculations to predict a value which denotes the probability of hostility of a given object in the area of interest.

The methodology revolves primarily around the calculation of the following parameters whose significance have been illustrated in Fig. 4–

- $D_T$, suspect-target distance.
- $D_{PN}$, suspect-potential destination distance.
- $I$, movement inefficiency index.
- Probability of hostility of suspect

### 2.2.1 Terminology Definitions –

Listed below are the definitions of a few terminologies which will be utilized to explain the approach.

- **Target**: The object or body which is vulnerable to attack and is to be protected.

- **Target Zone**: This is a zone of any shape or size, encompassing the target.

- **Target Zone Entry Point**: This is the point on the target zone boundary where the suspect makes its entry into the target zone.

- **Suspect-Target Distance**: It is the shortest distance between the suspect and the target at a given time.

- **Suspect-Potential Destination Distance**: It is the shortest distance between the suspect and a potential destination in the target zone for the suspect.

- **Suspect Zone**: This is a zone of any shape and size, encompassing the suspect.

- **Suspect Zone Entry Point**: It is a point on the boundary of the suspect zone where the suspect makes an entry into the suspect zone.

- **Movement Inefficiency Index**: It is the ratio between the actual distance travelled by the suspect from the suspect zone entry point to its current location and the shortest distance between the suspect's current location and the suspect zone entry point of the suspect.

- **Probability of Hostility**: It is the probability that a suspect is going to attack, abduct, sabotage, or steal the target or something contained within the target.

### 2.2.2 Method Outline –

Enumerated below is an outline of the methodology followed by this approach to quantify the measure of hostility of a given suspect.

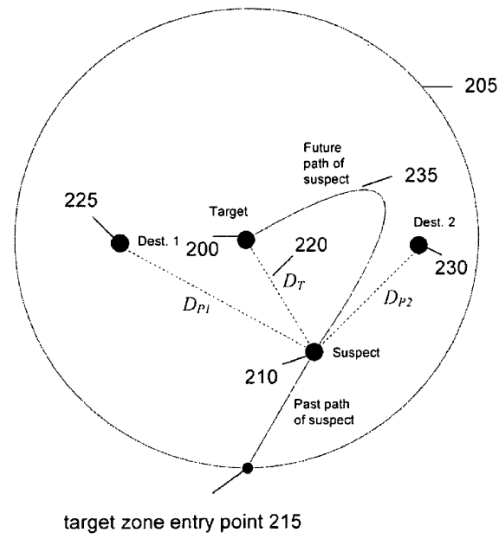

Fig. 4 Schematic diagram of target surrounded by a target zone and suspect inside the target zone [2]

Fig. 4 illustratively describes the significance of some of the terminologies described in Sec. 2.2.1. The circular shape of the target zone is for illustrative purposes only and may pragmatically be of any shape or size as long as it conforms to the definition of the target zone. The numbers illustrated correspond to the diagrammatic structures they are linked with.

Fig. 4 basically epitomizes the workings of the method as it does in different situations. Here, we have a single target (200) which is to be protected, a target zone (205) constructed around it and a single suspect (210). The suspect enters the target zone at the target zone entry point (215). Now since the suspect has entered the target zone, it will be subject to probation for the determination of its probability of hostility to the target. Henceforth, the location of the suspect is constantly mapped from the time it entered the target zone, and its probability of hostility is continuously calculated. Fig. 4 is a provisional snapshot of an intermediate situation after the suspect has entered the target zone so as to give an insight into the workings of the process, the flowchart of which, is given below –

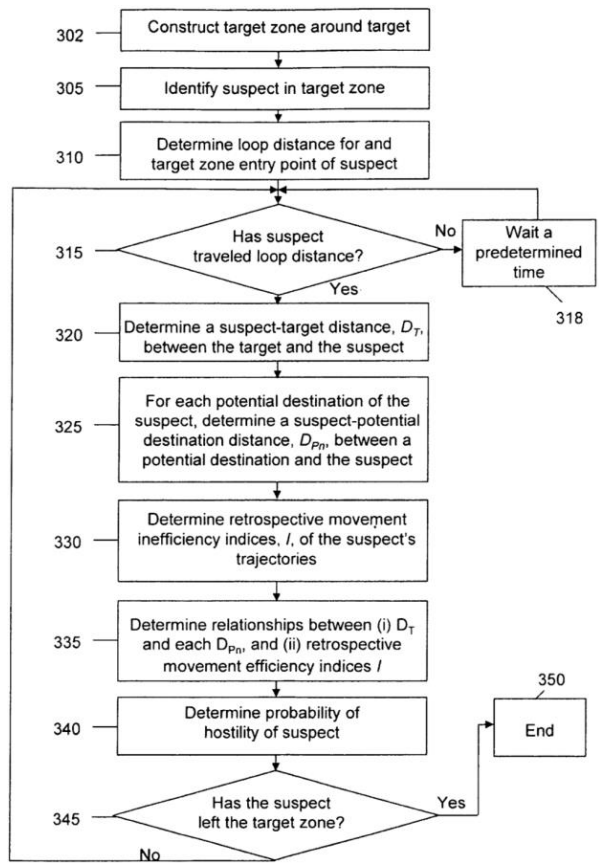

Fig. 5 Methodology flowchart of the system workings [2]

The flowchart depicted in Fig. 5. briefly illustrates and outlines the procedures involved. Therefore, as we can clearly see, this method keeps calculating the hostility probability values of the suspect till the time it is inside the target zone. It is therefore upto the user observing the values to set a threshold limit on the hostility values which when crossed, shall elicit initiation of defensive measures to counter the suspect. The method although, is purely analytical and deterministic and is basically built around some mathematical assumptions pertaining to preconceived notions about how hostile objects behave or are supposed to behave. It therefore, cannot encompass new emerging attack/defence trends and tactics.

## 3. Our Solution

The shortcomings of the aforementioned approaches elicit the need for a system which is fully automated, accurate, and is capable of keeping up with the constantly evolving field of attack/defence tactics. A perfect solution should be able to somehow incorporate the human characteristic of *intuition* into its functionality because this is the very characteristic which causes humans to make very accurate decisions when it comes to identifying hostile behavior. To make for the shortcomings, we have devised a solution which is completely automatic, has been found to be accurate, and has the ability to *learn* or adapt itself to new trends in attack/defence techniques. This has been achieved by the deployment of artificial neural networks. The inherent quality of fault tolerance of artificial neural networks makes them an excellent choice for being used to solve problems having an ambiguous nature such as this. Moreover, the ability of a neural network to learn arms it with self-adapting capabilities.

Neural networks, with their ability to derive correlations from complex and imprecise data, are utilized here to identify patterns in the movements of the objects in the area of interest. Out of the two methods of training a neural network namely *supervised* and *unsupervised*, we will be utilizing both learning techniques to train our neural networks specialized to this domain. The system so proposed, has two functional objectives at the highest level of abstraction –

- Object Tagging
- Object Hostility Classification

The object tagging module utilizes unsupervised learning techniques to achieve its objective, whereas the object hostility classification module is set to utilize supervised learning techniques.

## 3.1 Object Tagging

Given the input feed, it is imperative to uniquely identify each of the vessels and remember their identities so as to avoid mistaking one vessel to be another.

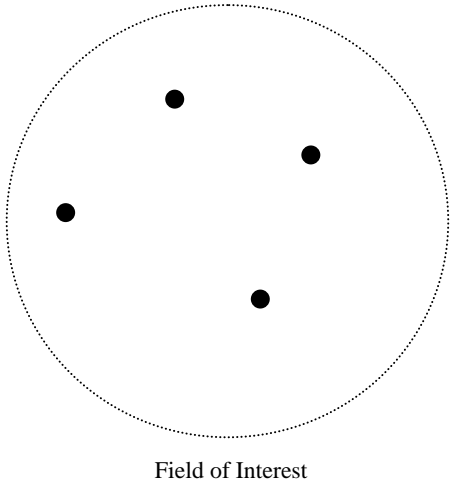

Field of Interest

Fig. 6 Objects in a field of interest, with no identification.

As is evident in Fig. 6, the blips shown in the figure are just locations of different objects in the area of interest, with no identity of its own. It is imperative to distinguish one object from the other for the working of this system. Hence, each object shall be uniquely *tagged* and assigned an ID number. A table called *location table* would be maintained in the memory which would map the location of each object to its ID number.

Table 1: *Location Table* mapping object location to ID number

| Object ID | Coordinate 1 | Coordinate 2 |
|-----------|--------------|--------------|
| 001       | 124          | 256          |
| 046       | 056          | 914          |
| 012       | 451          | 652          |
| 146       | 104          | 652          |
| 005       | 743          | 016          |

Table 1 illustrates as to how the records of the ID number-location mappings of each object may be maintained in memory. Here, Coordinate 1 refers to one of the coordinate parameters (such as X coordinate) and Coordinate 2 to the other.

For achieving this, Self-Organizing Maps (SOM) are deployed, as they have the ability to map a higher dimensional input space onto a lower dimensional space (for easy computation). The neuron topology to be deployed for this application with a 2-D input space will preferably be 2-Dimensional. In operation, as the radar feed starts coming, the radar inputs will be projected on a completely untrained SOM, which is expected to take a few inputs from the same before it starts to effectively uniquely identify the moving objects.

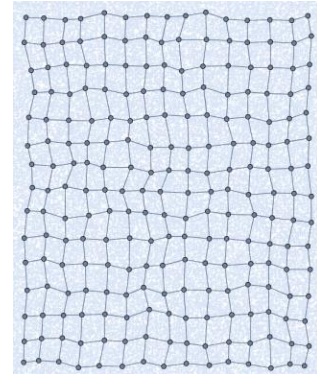

Fig. 7 An untrained self-organizing map (SOM) [3].

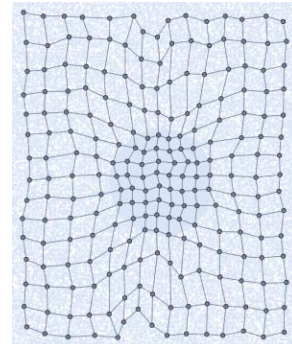

Fig. 8 A partially trained self-organizing map (SOM) [3].

Fig. 7 denotes the initial configuration of a SOM with no training and Fig. 8 illustrates how the SOM changes when data with only one object at the center is used to train the SOM. Similarly, other objects in the field of observation are tagged.

## 3.2 Object Hostility Classification

The second and final module of the system would have the objective to classify each of the uniquely identified vessels, as enumerated by the previous module as hostile or non-hostile based on the movement patterns. This module would be deployed as another artificial neural network which would undergo supervised training to develop its own notions of hostile object behavior characteristics. A two layer feed-forward network will be deployed in our application. The activation function for the neurons will be the logistic function defined as:

$$out = f_{sig}(net) = \frac{1}{1 + e^{-net}} \quad (1)$$

The number of inputs to the neural network will be the number of attributes pertaining to each object multiplied by the maximum number of objects in the area of interest. The number of outputs will simply be the number of objects in the area of interest with each output corresponding to each of the objects in the area of interest. The value of the output neurons will range from 0 to 1 as it shall denote the probability of hostility of the object.

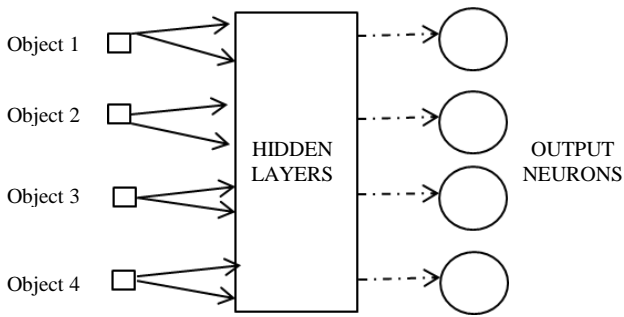

Fig. 9 Neural Network structure to be utilized for object hostility classification.

## 3.3 Process Flowchart

A brief outline of the entire methodology followed by this system is illustrated by the flowchart given below -

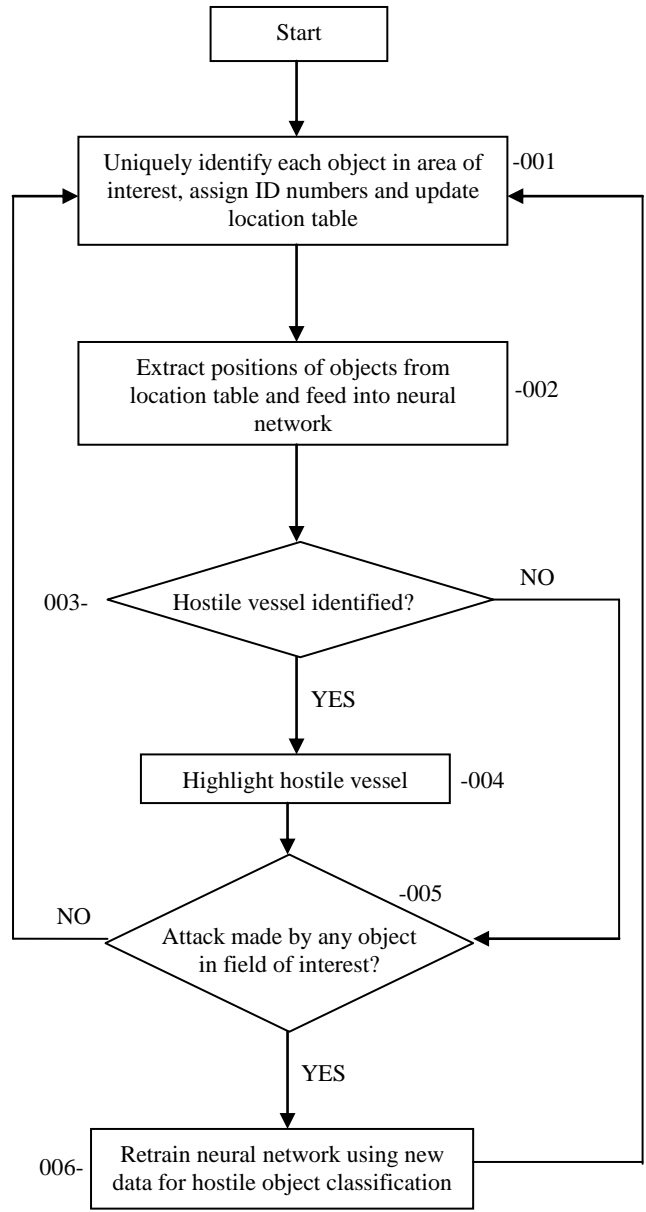

Fig. 10 Process Flowchart of Hostile Object Detection Technique proposed.

As shown in Fig. 10, firstly the objects in the area of interest are uniquely identified and tagged (001). The location table as defined in Sec. 3.1 is created and initialized. Subsequently, the locations corresponding to each of the uniquely identified objects are extracted from the location table and fed into the neural network (002). If any hostile vessel is identified to be hostile by the feed forward neural network, it is highlighted (004). Then, it is checked if any object in the field of interest has performed an act of hostility; so if the system had failed to highlight the hostile object beforehand (at step 004), the system retrains itself so as to prevent a similar prediction failure in the future (006). Therefore, step 006 explicates the self-learning capabilities of this system; i.e. in the event of a failure, the system *learns* from it and accordingly improves itself to maximize the chances of its success in the future.

## 4. Conclusions

From the discussion pertaining to the popular approaches elucidated in this paper, we can agree that an analytical approach is not a comprehensive solution. As previously stated, it is imperative to incorporate the human element of *intuition*. Therefore, a solution which tries to mimic the workings of human intuition has been outlined in this paper as a fully automated approach. The solution posited herein, covers the weaknesses of the current approaches and may be further modified or extended to encompass a multitude of other factors. The solution so proposed, seeks to serve as an outline for a general framework which may be specialized according to domain of application.

## 5. Future Work

This framework finds a promising application in the domain of cyber security and network packet analysis, where it may be deployed to sniff out malicious packets of data. The system may also be coupled with maritime surveillance hardware such as radars installed on sea ports to monitor sea traffic and look for potential threats.

**Souham Biswas** is a B.Tech. 4[th] year Computer Science & Engineering student at J.K. Institute of Applied Physics & Technology, University of Allahabad, Allahabad. He has achieved various national level accolades in the fields of embedded systems and application of artificial intelligence in embedded systems from IIT Bombay in the years 2013 and 2014 consecutively. He is also an avid software developer with his expertise spanning across domains of cloud computing and database management to scientific computing and machine learning. His areas of interest are Machine Learning, Artificial Neural Networks, Embedded Systems, Robotics and Software Design.

**Manisha J. Nene** is the Assistant Professor at the Dept. of Computer Engineering at Defence Institute of Advanced Technology, Pune under Defence Research & Development Organization, Ministry of Defence, Govt. of India.